\documentclass[10pt]{article}
\usepackage{eurosym}
\usepackage{amssymb}
\usepackage{amsmath}
\usepackage{subcaption}
\usepackage{graphicx}
\usepackage{cite}
\usepackage{geometry}
\usepackage{color}
\begin{document}

\title{Thermodynamics of Schwarzschild black hole surrounded by quintessence
with generalized uncertainty principle}
\author{B. C. L%
\"{u}tf\"{u}o\u{g}lu\thanks{%
bclutfuoglu@akdeniz.edu.tr (corresponding author)} \\
Department of Physics, Akdeniz University, Campus 07058, Antalya, Turkey, \\
Department of Physics, University of Hradec Kr\'{a}lov\'{e}, \\
Rokitansk\'{e}ho 62, 500 03 Hradec Kr\'{a}lov\'{e}, Czechia.\ \and B. Hamil \thanks{%
hamilbilel@gmail.com} \\
D\'{e}partement de TC de SNV, Universit\'{e} Hassiba Benbouali, Chlef,
Algeria. \and L. Dahbi\thanks{%
l.dahbi@ens-setif.dz} \\
Teacher Education College of Setif, Messaoud Zeghar, Algeria.}
\date{}
\maketitle

\begin{abstract}
In this manuscript, we consider a deformation on the Heisenberg algebra and investigate the effects on the thermodynamics of the Schwarzschild black hole that is surrounded by quintessence matter. To this end, we obtain the temperature, entropy, and heat capacity functions of the black hole by using the standard laws of thermodynamic according to the considered deformation. We show that upper and lower bound values appear on these functions based on the quintessence and deformed algebra. Then, we derive the corrected density of quintessence matter and the black hole's equation of state functions. We compare these results with the standard Schwarzschild black hole with and without quintessence with the graphical methods and interpret the quantum deformation effects.
\end{abstract}

\begin{description}
\item[PACS numbers:] 04.70.Dy, 04.70.-s

\item[Keywords:] Black hole thermodynamics, Quintessence, GUP formalism.
\end{description}

\section{Introduction}
Black holes are probably the most mysterious objects in the universe. In recent years, we  have observed an increase in the studies which {examine} black hole physics and thermodynamics in the modern cosmology literature \cite{Caldarelli2000, Page2005,Jacobson2010, Zhang2015,Appels2016,Astorino2017,Dehghani2018, Bayraktar2018,Hu2019, Yao2019, Krtous2020, Sharif2020, Gregory2021, Garcia2021, Sinha2021,Molina2021}. The basic assumption in these studies is that{:} "black holes {could} be taken as a thermal system, so the well-known thermodynamics laws could be used to interpret their nature". Such a correlation is originally presented by Hawking and Bekenstein a half-century ago \cite{H1, H2, B1}. According to them, black holes { are thermodynamic} objects that emit radiation from their event horizon with a characteristic temperature associated with their surface gravity. { Furthermore, the entropy functions of these objects are linearly proportional to the event horizon areas in Planck units,} and regarding the second law of thermodynamics their surface areas do not decrease. { With these revolutionist propositions,} they examined    {some properties} of the black holes, such as Hawking temperature and mass functions, within the framework of statistical mechanics and they concluded that the first law of thermodynamics was not violated  \cite{2}. 

The recent reliable observational astronomical {evidence} shows that the universe expands at an accelerating rate \cite{Riess1,Riess2,Perlmutter1999}.  To explain this acceleration, the presence of dark energy is postulated.  {In} the current model, dark energy is { considered} as another form of energy with negative pressure, and is {estimated to exist}  approximately {in seventy  percent}  of  the  total  energy density of the universe. { The simplest way to model this acceleration mathematically is given by the definition of a} cosmological constant \cite{Padman2003}. {However,} its experimental value is much smaller than its {theoretically expected} value \cite{SW1989}, {therefore,} alternative models have been put forward. These models are based on dynamic scalar fields, and they differ from each other by a parameter value which indicates the ratio of pressure to the energy density of the dark energy \cite{carroll1998, Khoury2004, Picon2000, Padman2002,Caldwell2002,Gasperini2002}.
For a detailed discussion and comparison among these models, {we refer} the readers to have a look at the review given in  \cite{Copeland2006}.

A dark energy surrounding the black hole is thought to have important effects on black hole thermodynamics. There are dozens of studies examining this scenario in the literature. For example,  in 2003, Kiselev took { the quintessence matter, which is among the black energy candidates, into account} and investigated the Schwarzschild black hole thermodynamics, assuming that it surrounds the black hole \cite{Kiselev}. { In 2008, Chen \emph{et al.} discussed the Hawking radiation in a $d$-dimensional spherically symmetric static  black hole with the presence of quintessence matter \cite{Chen2008}. }Later, Wei and Chu explored the thermodynamics properties of the Reissner - N\"ordstrom (NR) black hole surrounded by quintessence matter
\cite{Wei2011}. In 2013, Wei, this time with Ren, studied the thermodynamics of the NR black hole in de - Sitter quintessence spacetime \cite{Wei2013}. The same year, Fernando examined Narai type black holes with quintessence matter \cite{Fernanado2013a, Fernanado2013b}. Three years later, in a series of work Ghaderi and Malakolkalami presented the effects of quintessence matter on the thermodynamic functions  of the Bardeen \cite{Ghaderi2016a}, Schwarzschild, and RN black holes \cite{Ghaderi2016b}.
In 2019, Shahjalal considered some quantum correction to the Schwarzschild metric and compared the thermodynamics of the black hole with the presence and absence of the quintessence matter \cite{3}. Last year Haldar and Biswas showed that the thermodynamic volume of Bardeen anti de Sitter black hole is not equal to the geometric volume in the presence of quintessence matter \cite{Haldar2020}. Very recently, Ndogmo \emph{et al.} computed various thermodynamic quantities of a rotating non-linear magnetic-charged black hole embedded in the quintessence matter and discussed the phase transition \cite{Ndogmo2021}.

On the other hand, it is a very well known fact that quantum physics deals with the physical properties of nature at the atomic scale which classical physics cannot. Quantum mechanics is the mathematical toolbox that is used to explain the aspects of nature at atomic and subatomic scales. One of the mystery of this formalism is the operator concept. According to Heisenberg algebra, the operation order of the position and momentum operators on the wavefunction is important and { it} differs from each other up to a constant factor. If one takes the quantum theory of gravity into account, the Heisenberg algebra has to be extended with quantum gravitational corrections which lead the generalization of the Heisenberg algebra. As a natural consequence of this, the Heisenberg uncertainty principle (HUP), which is a measure of the observable values of the position and momentum operators, { must be replaced by} the generalized (gravitational) uncertainty principle (GUP). It is worth noting that such an extension is not unique, therefore different GUP {predictions  exist}. Each { of the non-equivalent algebric deformations}  produces different new features, such as measurable minimal length, maximum momentum, etc.. \cite{10,11}

In the literature, we observe that the GUP is frequently employed in the papers that examine black hole thermodynamics \cite{Adler2001, Nouicer2007, Myung2007,  Gangopadhyay2014, Buon0, Hassan2019a, Buon1, Hassan2020a, Hassan2021a,Bilel1, Bilel2, Bilel3,  Petruzziello2021, new}. However, to the best of our knowledge,  no one has studied the black hole thermodynamics surrounded by the quintessence matter within the framework of the deformed Heisenberg algebra. Since the GUP formalism introduces new features, such as the existence of a minimal length  and a maximal observable momentum values, we believe that this novelty could play important role on the interpretation of the thermodynamics of black holes surrounded by the quintessence matter.

With this motivation we prepare the  manuscript by using natural units.  The outline is as follows: In Sec. \ref{sec:sec2} we introduce the metric of the considered black hole {surrounded by the} quintessence matter. Then, in Sec. \ref{sec:sec3}, we briefly describe the deformed Heisenberg algebra, and derive the temperature, heat capacity and entropy functions, respectively. After we examine these thermodynamic functions in three particular {values} of the quintessence state-parameter, we obtain the GUP-corrected energy matter density and equation of state functions. We analyze the findings and conclude the manuscript in the final section.

\section{Schwarzschild black hole surrounded by quintessence}\label{sec:sec2}

We start by considering the Kiselev's paper \cite{Kiselev},  which is based on the derivation of the exact solution of Einstein equations  of a static spherically symmetric  black hole that is surrounded by the quintessence energy-matter. In that work, Kiselev expressed the general form of metric to be in the form of
\begin{equation}
ds^{2}=-N\left( r\right) dt^{2}+\frac{1}{N\left( r\right) }%
dr^{2}+r^{2}d\Omega _{2}^{2},
\end{equation}%
where for the Schwarzschild black hole, $N\left( r\right) $ can be taken as%
\begin{equation}
N\left( r\right) =1-\frac{2M}{r}-\frac{{\alpha_{\omega_q} }}{r^{3\omega _{q}+1}}.
\end{equation}%
Here, $M$ denotes the mass of the black hole, $\omega _{q}$ is the quintessential state parameter, and  ${\alpha_{\omega_q}} $ is the positive normalization factor that depends on the density of quintessence matter {(For simplicity, hereafter,  we will denote it without the subscript)}. In the range of $-1<\omega _{q}<-1/3$, the quintessence successfully explains the accelerated expansion of the universe. Note that, when $\alpha $  is taken to be equal to zero, the metric reduces to the ordinary Schwarzschild black hole metric.  

Kiselev also showed that in this quintessence matter model the energy-momentum tensor has the following components
\begin{eqnarray}
&&T_{t}^{t}=T_{r}^{r}=-\rho , \\
&&T_{\theta }^{\theta }=T_{\phi }^{\phi }=\frac{\rho }{2}\left( 3\omega
_{q}+1\right),
\end{eqnarray}%
where the pressure, $P_{q}$, and the matter-energy density, $\rho _{q}$, are written in the form of%
\begin{eqnarray}
P_{q}&=&\omega _{q}\rho _{q}, \label{Pressure} \\
\rho _{q}&=&-\frac{3}{2}\frac{\alpha \omega _{q}}{r^{3\left( \omega_{q}+1\right) }}. \label{energymatter}
\end{eqnarray}%
If {we consider} the quintessence state parameter and normalization factor values {that are given above, then we } can conclude that the matter-energy density takes only positive values, while the pressure of the quintessence matter {takes only} negative values.  

\begin{figure}[tbph]
\centering
\includegraphics[scale=1]{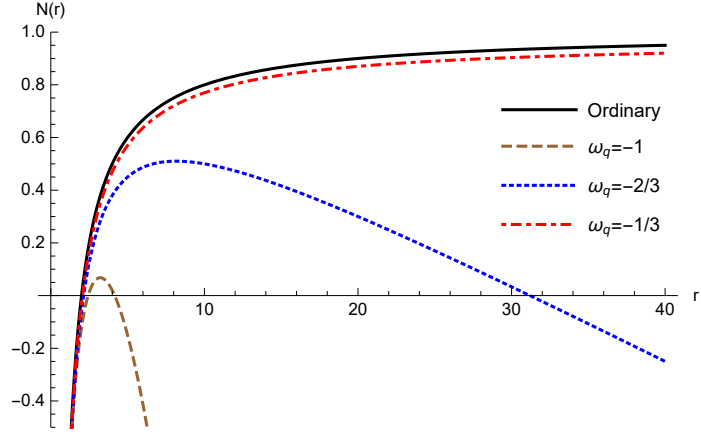}
\caption{$N\left( r\right) $ versus $r$ for the particular values of the quintessence state parameter.} \label{fig1}
\end{figure}

After the introduction of these facts, we intend to clarify the event horizon radius. In order to determine them, we set the following condition 
\begin{equation}
\left. N\left( r\right) \right \vert _{r=r_{H}}=0,
\end{equation}%
which is equivalent to solving %
\begin{equation}
\left. 1-\frac{2M}{r}-\frac{\alpha }{r^{3\omega _{q}+1}}\right \vert 
_{r=r_{H}}=0. \label{eq7}
\end{equation}
We classify the solutions in three particular cases:
\begin{itemize}
\item \bigskip For $\omega _{q}=-2/3$, two event horizon radii appear, namely \emph{inner}, $r_{in}$, and \emph{outer}, $r_{out}$, as follows: 
\begin{eqnarray}
r_{in}&=&\frac{1-\sqrt{1-8M\alpha }}{2\alpha }, \label{-23rin} \\
r_{out}&=&\frac{1+\sqrt{1-8M\alpha }}{2\alpha }.
\end{eqnarray}
It is worth noting that the outer horizon is commonly called as the quintessence horizon, alike to the cosmological horizon in the de-Sitter spacetime \cite{2,3,4,5}.
\item \bigskip For $\omega _{q}=-1/3$, only one horizon arises
\begin{equation}
r_{H}=\frac{2M}{1-\alpha }.
\end{equation}
\item \bigskip In the case of $\omega _{q}=-1$,  we obtain the de Sitter-Schwarzschild solution if we take 
\begin{equation}
\alpha =\frac{\Lambda }{3}.
\end{equation}%
\end{itemize}
Before ending this section, we present a graphical demonstration of these cases. We take $M=1$ and $\alpha =0.03$, and we plot $N(r)$ versus distance in Fig. \ref{fig1}.

We observe that,  $\omega_q=-1/3$ case mimics the ordinary case, since the event horizon radius in each cases differ by themselves just by a constant that is extremely small. In the case of $\omega_q=-2/3$, real-valued inner and outer event horizon radii appear since $\alpha \leq 1/(8M)$ condition is satisfied. Moreover, with this particular choice of parameters two horizon occur in the $\omega_q=-1$ case.

\section{Thermodynamic features of the black hole} \label{sec:sec3}

At the energies near to the Planck scale, the conventional concepts of time and space break down. Considering the effect of gravity, the existence of a minimal length scale becomes necessary. However, such a minimal length does not exist in the usual Heisenberg algebra. Therefore, one can take the GUP instead of the HUP into account which is given with the natural units in the form of \cite{10,11}:
\begin{eqnarray}
\left( \Delta P\right) \left( \Delta X\right) \geq \frac{1}{2}\left( 1+\beta\left( \Delta P\right) ^{2}\right) ,  \label{1}
\end{eqnarray}%
where $\beta $ is a small non-negative deformation parameter that is proportional to the Planck length, and it is defined within the generalized Heisenberg algebra of the form of $[x,p]=i(1+\beta p^2)$. This deformation leads to a minimum uncertainty in the position \cite{10,11}.
\begin{equation}
\left( \Delta X\right) _{\min }=  \sqrt{\beta }.
\end{equation}%
{ It's worth noting that there are other scenarios where the deformation parameter is taken as a negative quantity \cite{B01,B02,B03,B04}. Moreover,  there are very interesting works in which the deformation parameter is regarded as a dynamical variable within a more general perspective instead of being a constant \cite{B05,B06}. After this notice, we}
start by solving Eq. (\ref{1}) with respect to $\left( \Delta P\right) $. We find
\begin{equation}
\frac{\left( \Delta X\right) }{\beta }\left( 1-\sqrt{1-\frac{\beta  }{\left(
\Delta X\right) ^{2}}}\right) \leqslant \left( \Delta P\right) \leqslant 
\frac{\left( \Delta X\right) }{\beta  }\left( 1+\sqrt{1-\frac{\beta  }{\left(
\Delta X\right) ^{2}}}\right) .  \label{3}
\end{equation}%
We observe that for $\frac{\beta  }{\left( \Delta X\right) ^{2}}<<1$, the left hand side of the
inequality produces some small corrections to the HUP, while the right hand side implies an upper bound value to the momentum uncertainty, therefore, $\left(\Delta P\right) $  cannot be arbitrarily large.

Hereafter, we study the effects of the GUP on the Schwarzschild black hole black hole
surrounded by quintessence. In the semiclassical case, if the entropy, $S$, is assumed to be a function of the black hole area, $A$, then, the temperature of the black hole can be expressed with a relation between them as given in \cite{Xiang}
\begin{equation}
T=\frac{\kappa }{8\pi }\frac{dA}{dS},  \label{T}
\end{equation}%
where $\kappa$ is the surface gravity at the outer horizon, and in our case it is equal to
\begin{equation}
\kappa =-\lim_{r\rightarrow r_{H}}\sqrt{-\frac{g^{11}}{g^{00}}}\frac{\left(
g^{00}\right) ^{\prime }}{g^{00}}=\frac{1}{r_{H}}\left( 1+\frac{3\alpha
\omega _{q}}{r_{H}^{3\omega _{q}+1}}\right).  \label{surface}
\end{equation}%
In \cite{Xiang},  it's shown that if a black hole absorbs a particle then its area changes proportionally with the particle's mass and size that are associated with the uncertainties of momentum and position. Furthermore, such a minimal change of the area leads to a  change in the entropy which can not be smaller than $\ln 2$. Therefore, one can write
\begin{equation}
\frac{dA}{dS}\simeq \frac{\left( \Delta A\right) _{\min }}{\left( \Delta
S\right) _{\min }}\simeq \frac{\epsilon }{\ln 2}\left( \Delta X\right)
\left( \Delta P\right) ,\end{equation}
where $\epsilon $ denotes the calibration factor. We use $\Delta X\simeq 2r_{H}$, and employ $\left( \Delta
P\right) $ from Eq. \eqref{3} to evaluate 
\begin{equation}
\frac{dA}{dS}\simeq \frac{4\epsilon r_{H}^{2}}{\beta \ln 2}\left( 1-\sqrt{1-%
\frac{\beta  }{4r_{H}^{2}}}\right) .  \label{area}
\end{equation}
Then, by substituting Eqs. \eqref{surface} and \eqref{area} in Eq. \eqref{T}, we obtain a relation between the temperature and horizon radius of the black hole in the form of
\begin{equation}
T=\frac{1}{2\pi }\frac{\epsilon r_{H}}{\beta  \ln 2}\left( 1+\frac{3\alpha
\omega _{q}}{r_{H}^{3\omega _{q}+1}}\right) \left( 1-\sqrt{1-\frac{\beta  }{%
4r_{H}^{2}}}\right). \label{Td}
\end{equation}%
In the absence of the quintessence and the GUP deformation, $\left(\alpha=\beta=0 \right)$,  Eq. (\ref{Td}) reduces to $T=
\frac{\epsilon }{16\pi r_{H}\ln 2}$, which should be equal to the   Hawking temperature , $T_{H}=\frac{1}{4\pi r_{H}}$,  \cite{Hawking,Bekenstein}. Therefore, we determine the calibration factor as $\epsilon =4\ln 2$. {Then}, we rewrite the GUP-corrected Hawking temperature as
\begin{equation}
T=\frac{2r_{H}}{\pi \beta }\left( 1+\frac{3\alpha \omega _{q}}{%
r_{H}^{3\omega _{q}+1}}\right) \left( 1-\sqrt{1-\frac{\beta }{4r_{H}^{2}}}%
\right) .  \label{TGUP}
\end{equation}
For $\beta =0$, Eq. \eqref{TGUP} reduces to the Hawking temperature of Schwarzschild
black hole surrounded by the quintessence in the HUP limit
\begin{equation}
T=\frac{1 }{4\pi   r_{H} }\left( 1+\frac{3\alpha \omega _{q}}{%
r_{H}^{3\omega _{q}+1}}\right) .  \label{TGUPa}
\end{equation}
Regarding the fact that the temperature of a black hole must be real-valued, we find a constraint on the radius that bounds the radius value from above.
\begin{eqnarray}
r_H \leq \sqrt[(3\omega_q+1)]{ -3 \alpha  \omega_q}. \label{rup}
\end{eqnarray}
It is worth noting that {in this study} $\alpha>0$ and $\omega_q<0$, and hence the horizon radius has a positive root value unless the quintessence state parameter is equal to $-1/3$.  {For} $\omega_q=-1/3$, the horizon radius is not bounded from above. Therefore, we conclude that the {greatest possible event horizon is} determined by the quintessence matter that surrounds the black hole.

\begin{figure}[htbp]
\centering
    \begin{subfigure}{0.45\linewidth}
\includegraphics[width=\linewidth]{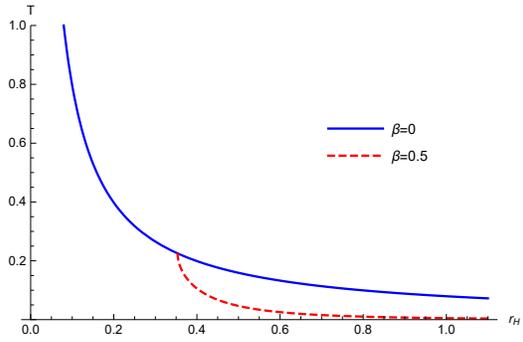} 
    \caption{HUP and GUP cases for $\alpha=0$.}
\label{fig:2a}
    \end{subfigure}\hfill
    \begin{subfigure}{0.45\linewidth}
\includegraphics[width=\linewidth]{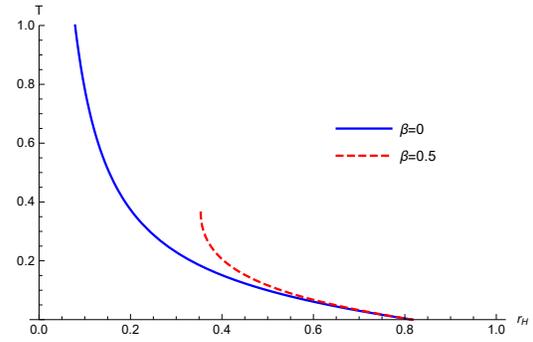}
    \caption{$w_q=-1$ case.}
\label{fig:2b}
    \end{subfigure}

    \begin{subfigure}{0.45\linewidth}
\includegraphics[width=\linewidth]{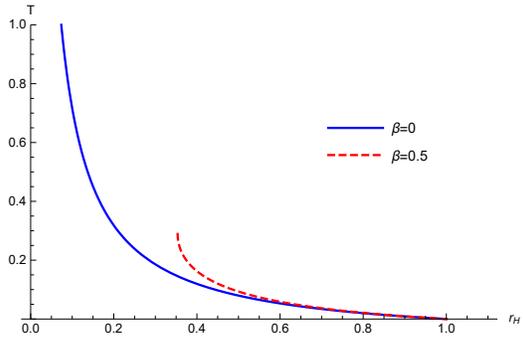}
    \caption{$w_q=-2/3$ case.}
\label{fig:2c}
    \end{subfigure}\hfill
    \begin{subfigure}{0.45\linewidth}
\includegraphics[width=\linewidth]{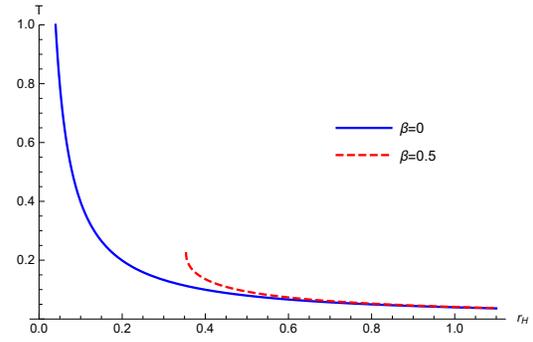}
    \caption{$w_q=-1/3$ case.}
\label{fig:2d}
    \end{subfigure}
\caption{Temperature versus horizon radius for $\alpha=0.5$.}
    \label{fig:2}
    \end{figure}

On the other hand, in the GUP-corrected case, according to Eq. \eqref{TGUP}, we observe that the modified Hawking temperature depends not only on the black hole's properties but also on the GUP parameter which leads to another constraint as
\begin{equation}
 r_{H}\geq \frac{ \sqrt{\beta }}{2}. \label{2a}
\end{equation}%
{This result indicates that the black hole's lowest radius value is based on the GUP.} Furthermore, in the GUP-corrected case a maximal temperature value {arises} in the form of
\begin{eqnarray}
T \leq \frac{1}{\pi \sqrt{\beta}}\left[ 1+3 \alpha \omega_q\left( \frac{\beta}{4}\right)^{-\frac{3\omega_q+1}{2}}\right].
\end{eqnarray}
In the absence of the quintessence matter, we find that the GUP-corrected temperature of the black hole can take values only in the following certain range
\begin{eqnarray}
0 < T\leq \frac{1}{ 2\pi \sqrt{\beta}}. \label{2b}
\end{eqnarray}
In Fig. \ref{fig:2} we depict the black hole temperature versus the horizon radius in the HUP and GUP approaches, respectively. In particular, in Fig. \ref{fig:2a}, we consider  the HUP and GUP cases in the absence of the quintessence matter. We observe that in the ordinary case there is no bound on the temperature and the radius. However, in the GUP case, the radius can take values after $0.35$ which corresponds the maximal temperature value $0.22$ as one can calculate from Eqs. \eqref{2a} and \eqref{2b}, respectively.   

We summarize all our findings in three particular cases:
\begin{itemize}
\item For $\omega _{q}=-1$, the GUP-corrected temperature of the quintessence surrounded Schwarzschild black hole can be calculated via
\begin{equation}
T_{(\omega _{q}=-1)}=\frac{2r_{H}}{\pi \beta }\left( 1-3\alpha
r_{H}^{2}\right) \left( 1-\sqrt{1-\frac{\beta }{4r_{H}^{2}}}\right) .
\end{equation}%
In this case the event horizon radius can only take values in the range of 
\begin{eqnarray}
\frac{1}{\sqrt{3\alpha }}\geq r_{H}\geq \frac{\sqrt{\beta}}{2},
\end{eqnarray}
and hence the temperature can vary in between 
\begin{eqnarray}
0 \leq T_{(\omega _{q}=-1)} \leq \frac{4-3 \alpha \beta}{4 \pi \sqrt{\beta}}. \label{tr1}
\end{eqnarray}
Note that, in the absence of the GUP,  there is no lower bound value of the radius, therefore there is no upper limit value for temperature. We depict this case in Fig. \ref{fig:2b}. We find that in the GUP case, the radius is confined in the range of $[0.35,0.82]$ while the temperature has a maximum value $0.37$.    

\item For $\omega _{q}=-2/3$, alike, the black hole temperature turns to the form of%
\begin{equation}
T_{(\omega _{q}=-2/3)}=\frac{2r_{H}}{\pi \beta  }\left( 1-2\alpha r_H\right)
\left( 1-\sqrt{1-\frac{\beta }{4r_{H}^{2}}}\right) ,
\end{equation}%
which limits the horizon radius to the following range 
\begin{eqnarray}
\frac{1}{2\alpha }\geq r_{H}\geq \frac{\sqrt{\beta }}{2}.
\end{eqnarray}
Therefore, in this case  the GUP-corrected temperature range becomes
\begin{eqnarray}
0 \leq T_{(\omega _{q}=-2/3)} \leq \frac{1- \alpha \sqrt{\beta}}{ \pi \sqrt{\beta}}. \label{tr2}
\end{eqnarray}
We illustrate this case in Fig. \ref{fig:2c}. We find that in the HUP and GUP cases, there is the same upper limit value, $1$, for the horizon radius. In the GUP case, the lower bound radius value, $0.35$, exists while in the HUP case does not. Moreover, in the GUP case, the temperature has a maximum value {of}
$0.29$.

\item For $\omega _{q}=-1/3$, Eq. \eqref{TGUP} becomes%
\begin{equation}
T_{(\omega _{q}=-1/3)}=\frac{2r_{H}}{\pi \beta }\left( 1-\alpha \right) \left(
1-\sqrt{1-\frac{\beta}{4r_{H}^{2}}}\right) .
\end{equation}
However, in this case, an upper bound on the horizon radius does not emerge. The GUP-corrected temperature varies in the range of
\begin{eqnarray}
0 < T_{(\omega _{q}=-1/3)} \leq \frac{1- \alpha}{ \pi \sqrt{\beta}}. \label{tr3}
\end{eqnarray}
We illustrate this case in Fig. \ref{fig:2d}. We observe that there is no upper bound on the horizon radii. However, in the GUP case, there is a lower radius value, $0.35${,} which corresponds a maximum temperature value of $0.23$. 
\end{itemize}

As a conclusion, we observe that all graphical demonstrations approve the predictions of our analyze. Moreover, these results are in a good agreement with the results given in \cite{Adler2001}.

\begin{figure}[htbp]
\centering
    \begin{subfigure}{0.45\linewidth}
\includegraphics[width=\linewidth]{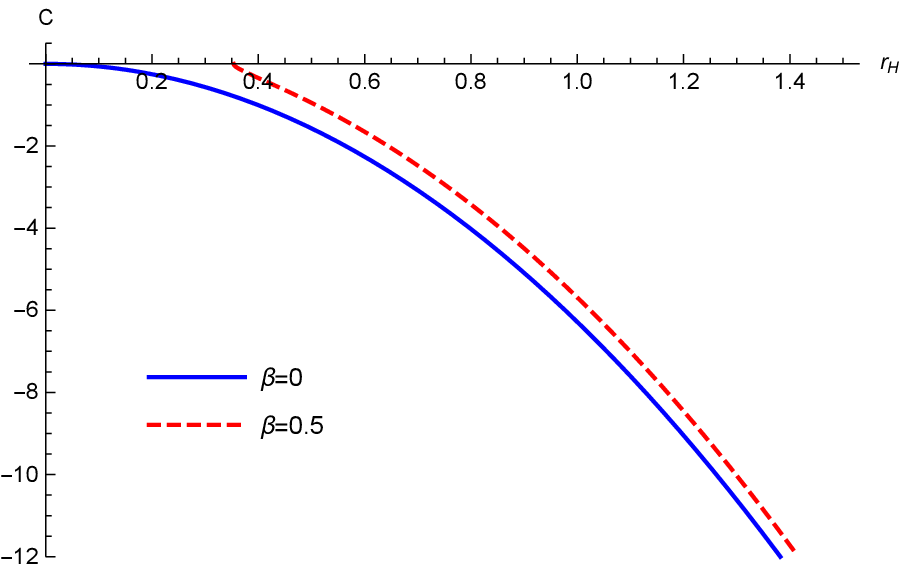} 
    \caption{HUP and GUP cases for $\alpha=0$.}
\label{fig:3a}
    \end{subfigure}\hfill
    \begin{subfigure}{0.45\linewidth}
\includegraphics[width=\linewidth]{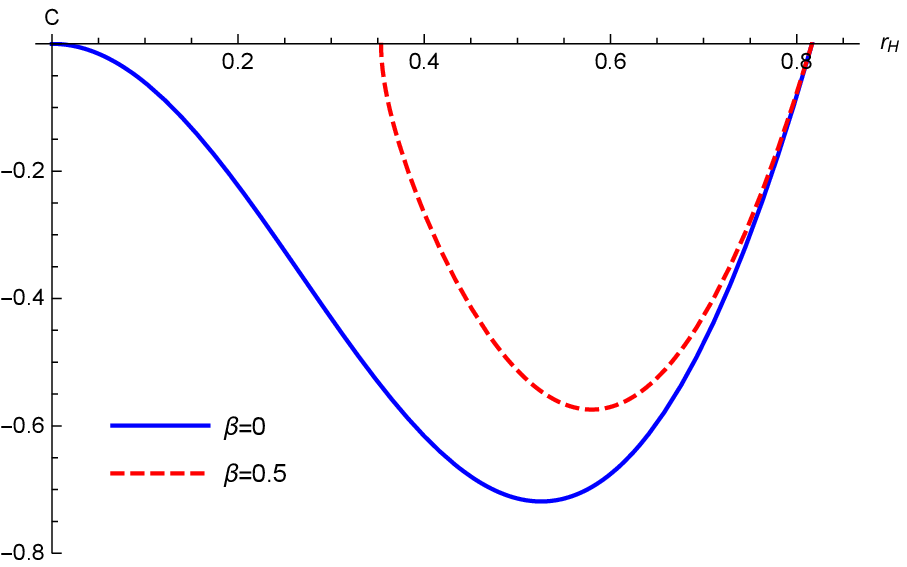}
    \caption{$w_q=-1$ case.}
\label{fig:3b}
    \end{subfigure}

    \begin{subfigure}{0.45\linewidth}
\includegraphics[width=\linewidth]{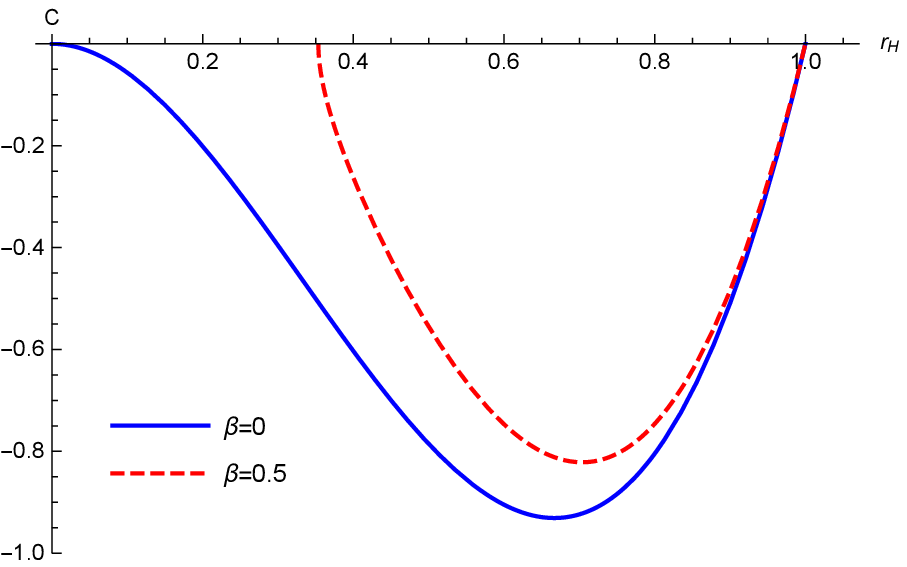}
    \caption{$w_q=-2/3$ case.}
\label{fig:3c}
    \end{subfigure}\hfill
    \begin{subfigure}{0.45\linewidth}
\includegraphics[width=\linewidth]{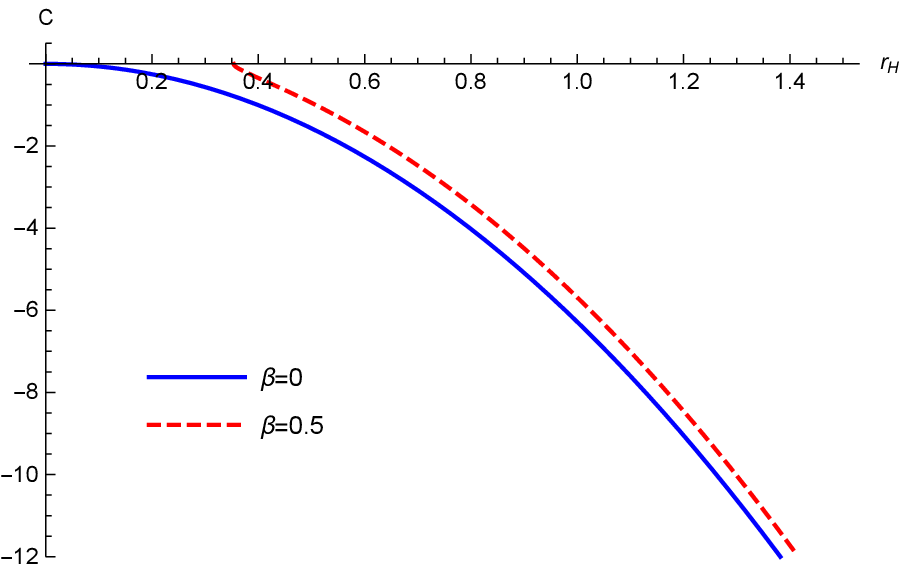}
    \caption{$w_q=-1/3$ case.}
\label{fig:3d}
    \end{subfigure}
\caption{Specific heat versus horizon radius for $\alpha=0.5$.}
    \label{fig:3}
    \end{figure}

Next, we derive the GUP-corrected heat capacity function of the black hole heat by employing the following thermodynamic relation
\begin{equation}
C=\frac{dM}{dT}.  \label{def}
\end{equation}
We find
\begin{equation}
C=-\frac{\pi \beta }{4}\frac{\left( 1+\frac{3\alpha \omega _{q}}{r_H^{3\omega
_{q}+1}}\right) \sqrt{1-\frac{\beta }{4r_H^{2}}}}{\left( 1-\frac{9\alpha
\omega _{q}^{2}}{r_H^{3\omega _{q}+1}}\right) \left( 1-\sqrt{1-\frac{\beta }{%
4r_H^{2}}}\right) +\frac{\beta }{4r_H^{2}}\frac{3\alpha \omega _{q}\left(
1+3\omega _{q}\right) }{r_H^{3\omega _{q}+1}}}. \label{SpHeatGUP}
\end{equation}
It is worth noting that when the heat capacity is equal to zero, then the black hole cannot exchange radiation with its surrounding space. This scenario is known as the black hole remnant. Regarding to this fact, we observe that  at
\begin{equation}
r_{rem}=\frac{\sqrt{\beta }}{2},  \label{rem}
\end{equation}%
Eq. \eqref{SpHeatGUP} tends to zero, so that black hole remnant exists. By substituting Eq. (\ref{rem}) in Eq. (\ref{TGUP}), we obtain the non-zero black hole remnant temperature
\begin{equation}
T_{rem}=\frac{1}{\pi \sqrt{\beta }}\left( 1+3\alpha \omega _{q}\left( \frac{2}{\sqrt{\beta }}\right) ^{3\omega _{q}+1}\right){,} \label{TremGUP}
\end{equation}
{with the ensuing mass
\begin{equation}
M_{rem}=\frac{\sqrt{\beta }}{4}\left( 1-\alpha \left( \frac{2}{\sqrt{\beta }}%
\right) ^{3\omega_{q}+1}\right) .  \label{a}
\end{equation}}
In the absence of quintessence matter, GUP-corrected specific heat function and the remnant temperature reduce to 
\begin{eqnarray}
C&=&-\frac{\pi \beta }{4}\frac{ \sqrt{1-\frac{\beta }{4r_H^{2}}}}{1-\sqrt{1-\frac{\beta }{%
4r_H^{2}}} }, \label{SpHeatGUPalphazero} \\
T_{rem}&=&\frac{1}{\pi \sqrt{\beta }}. \label{Tremalphazero}
\end{eqnarray}
Moreover, if we take $\beta=0$, we get the ordinary specific heat function  $C=-2\pi  r_H^2$ while a non-zero finite remnant temperature value does not exist. In Fig. \ref{fig:3}, we depict specific heat function versus the horizon radius. { Fig. \ref{fig:3a}  presents} the cases in the absence of the quintessence matter. We observe that the GUP-correction modifies this thermal feature at the small radius. We evaluate Eq. \eqref{Tremalphazero} and find the remnant temperature as $T_{rem}=0.45$. 

Next, we express the GUP-corrected specific heat and remnant temperature expressions for the particular quintessence state parameter. 
\begin{itemize}
\item For $\omega _{q}=-1$, Eqs. \eqref{SpHeatGUP}, \eqref{TremGUP}, {and \eqref{a}} reduce to
\begin{eqnarray}
C&=&-\frac{\pi \beta }{4}\frac{\left( 1-3\alpha r_H^2 \right) \sqrt{1-\frac{\beta }{4r_H^{2}}}}{\left( 1-9\alpha r_H^2 \right) \left( 1-\sqrt{1-\frac{\beta }{%
4r_H^{2}}}\right) +\frac{3\alpha \beta }{2}}, \label{SpHeatGUP1} \\
T_{rem}&=& \frac{1}{\pi \sqrt{\beta}}\left(1-\frac{3 \alpha \beta}{4} \right), \label{Trem-1}
\\
{M_{rem}}&=& {\frac{\sqrt{\beta }}{4}\left( 1-\frac{\alpha \beta }{4}\right).} \label{Mrem-1}
\end{eqnarray}

\item For $\omega _{q}=-2/3$, Eqs. \eqref{SpHeatGUP}, \eqref{TremGUP}, {and \eqref{a}} become%
\begin{eqnarray}
C=&-&\frac{\pi \beta }{4}\frac{\left( 1-2\alpha r_H \right) \sqrt{1-\frac{\beta }{4r_H^{2}}}}{\left( 1-4\alpha r_H \right) \left( 1-\sqrt{1-\frac{\beta }{%
4r_H^{2}}}\right) +\frac{\alpha \beta }{2r_H}}, \label{SpHeatGUP23} \\
T_{rem}&=& \frac{1-\alpha \sqrt{\beta}}{\pi \sqrt{\beta}}, \label{Trem-23}\\
{M_{rem}}&=& {\frac{\sqrt{\beta }}{4}\left( 1-\frac{\alpha \sqrt{\beta }}{2}\right).} \label{Mrem-23}
\end{eqnarray}

\item For $\omega _{q}=-1/3$, Eqs. \eqref{SpHeatGUP}, \eqref{TremGUP}, {and \eqref{a}} turn to the form of
\begin{eqnarray}
C&=&-\frac{\pi \beta }{4}\frac{ \sqrt{1-\frac{\beta }{4r_H^{2}}}}{1-\sqrt{1-\frac{\beta }{%
4r_H^{2}}} }, \label{SpHeatGUP13} \\
T_{rem}&=& \frac{1-\alpha}{\pi \sqrt{\beta}}, \label{Trem-13}\\
{M_{rem}}&=& {\frac{\left( 1-\alpha \right) \sqrt{\beta }}{4}.} \label{Mrem-13}
\end{eqnarray}
\end{itemize}

We depict the GUP-corrected specific heat functions for $\omega_q=-1$, $\omega_q=-2/3$ and $\omega_q=-1/3$ in Fig. \ref{fig:3b}, Fig. \ref{fig:3c} and
Fig. \ref{fig:3d}, respectively. For $\alpha=\beta=0.5$, we find the remnant temperatures as $0.37$, $0.29$ and $0.23$ for these cases. We note that these values are the same with the ones given in Eqs. \eqref{tr1}, \eqref{tr2} and \eqref{tr3}.

Next, we derive the entropy function of the black hole by employing the customary definition, that is given in the form of:
\begin{equation}
S=\int \frac{dM}{T}. \label{ent}
\end{equation}%
Using Eqs. \eqref{eq7} and \eqref{TGUP} in Eq. \eqref{ent}, we find  
\begin{equation}
S=\frac{\pi }{2}r_{H}^{2}\left( 1+\sqrt{1-\frac{\beta }{4r_{H}^{2}}}\right) -%
\frac{\pi \beta }{8}\ln r_{H}-\frac{\pi \beta }{8}\ln \left( 1+\sqrt{1-\frac{%
\beta }{4r_{H}^{2}}}\right).
\end{equation}%
We observe that the quintessence that surrounds the black hole does not affect the black hole's entropy. Instead the GUP correction
modifies the entropy. In the HUP limit, the entropy reduces to the usual expression, $S=\pi r_H^2=\frac{A}{4}$.

\begin{figure}[htbp]
\centering
    \begin{subfigure}{0.45\linewidth}
\includegraphics[width=\linewidth]{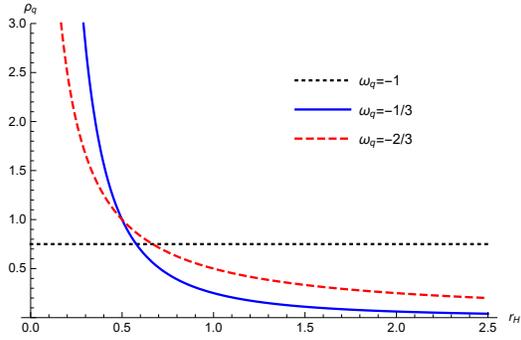} 
    \caption{HUP case for $\alpha=0.5$ and $\beta=0$.}
\label{fig:4a}
    \end{subfigure}\hfill
    \begin{subfigure}{0.45\linewidth}
\includegraphics[width=\linewidth]{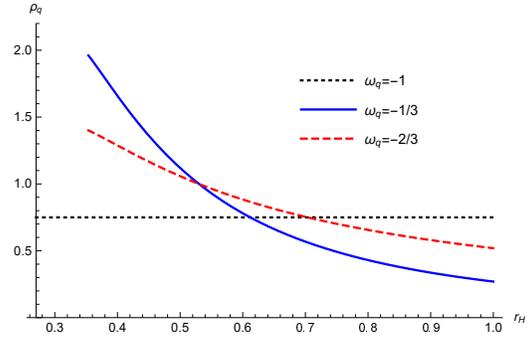}
    \caption{GUP case for $\alpha=\beta=0.5$.}
\label{fig:4b}
    \end{subfigure}
    \begin{subfigure}{0.45\linewidth}
\includegraphics[width=\linewidth]{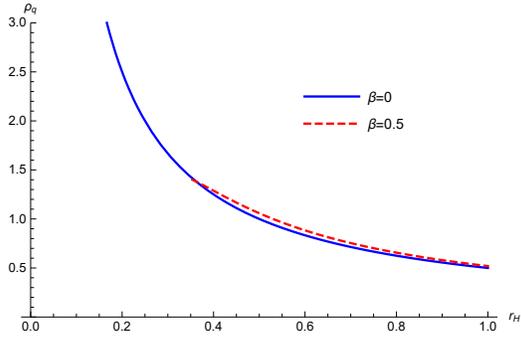}
    \caption{$w_q=-2/3$ case for $\alpha=0.5$.}
\label{fig:4c}
    \end{subfigure}\hfill
    \begin{subfigure}{0.45\linewidth}
\includegraphics[width=\linewidth]{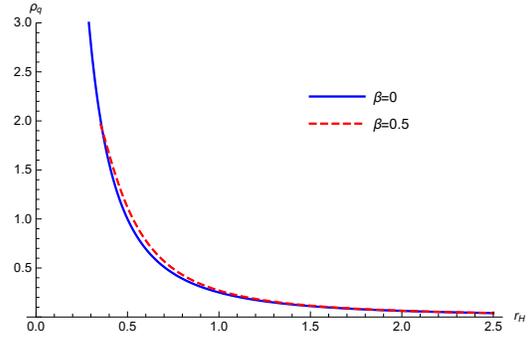}
    \caption{$w_q=-1/3$ case for $\alpha=0.5$.}
\label{fig:4d}
    \end{subfigure}
\caption{Energy-matter density of quintessence versus horizon radius. }
    \label{fig:4}
    \end{figure}

Then, we express the GUP-corrected energy-matter density of the quintessence by using the definition given in Eq. \eqref{energymatter}. We get 
\begin{eqnarray}
\rho _{q}&=& -\frac{3 {\alpha_{\omega _{q}}} \omega _{q}}{2}\left[ \frac{r_{H}^{2}}{2}%
\left( 1+\sqrt{1-\frac{\beta }{4r_{H}^{2}}}\right) -\frac{\beta }{8}\ln
r_{H}-\frac{\beta }{8}\ln \left( 1+\sqrt{1-\frac{\beta }{4r_{H}^{2}}}\right) %
\right] ^{-\frac{3\omega _{q}+3}{2}}, \label{rhoGUP}
\end{eqnarray}
which reduces to 
\begin{equation}
\rho _{q}=-\frac{3{\alpha_{\omega _{q}}} \omega _{q}}{2r_{H}^{3\omega _{q}+3}}. \label{rhoHUP}
\end{equation}%
for $\beta =0$. In the particular values of the quintessence state parameters the energy-matter density function reduces to the following forms:
\begin{itemize}

\item  For $\omega _{q}=-1$, it becomes the same constant that depends only on the normalization factor.
    \begin{eqnarray}
    \rho_q=\frac{3{\alpha_{(\omega _{q}=-1)}}}{2}.
    \end{eqnarray}
    
\item  For $\omega _{q}=-2/3$, GUP-corrected energy density function becomes
    \begin{eqnarray}
    \rho_q={\alpha_{(\omega _{q}=-2/3)}} \left[ \frac{r_{H}^{2}}{2}%
\left( 1+\sqrt{1-\frac{\beta }{4r_{H}^{2}}}\right) -\frac{\beta }{8}\ln
r_{H}-\frac{\beta }{8}\ln \left( 1+\sqrt{1-\frac{\beta }{4r_{H}^{2}}}\right) %
\right] ^{-\frac{1}{2}}, \label{rhoGUP-23}
    \end{eqnarray}
while in the HUP case it reduces to
\begin{equation}
\rho _{q}=\frac{{\alpha_{(\omega _{q}=-2/3)}}}{r_{H}}. \label{rhoHUP-23}
\end{equation}%
  
\item  For $\omega _{q}=-1/3$, GUP-corrected energy density function appears as 
    \begin{eqnarray}
    \rho_q=\frac{{\alpha_{(\omega _{q}=-1/3)}}}{2} \left[ \frac{r_{H}^{2}}{2}%
\left( 1+\sqrt{1-\frac{\beta }{4r_{H}^{2}}}\right) -\frac{\beta }{8}\ln
r_{H}-\frac{\beta }{8}\ln \left( 1+\sqrt{1-\frac{\beta }{4r_{H}^{2}}}\right) %
\right] ^{-1}, \label{rhoGUP-13}
    \end{eqnarray}
while in the HUP case it turns to
\begin{equation}
\rho _{q}=\frac{{\alpha_{(\omega _{q}=-1/3)}}}{2r_{H}^2}. \label{rhoHUP-13}
\end{equation}%
\end{itemize}
We present the the behavior of the energy-matter density of quintessence in Fig. \ref{fig:4}. In Fig. \ref{fig:4a} and Fig. \ref{fig:4b} we illustrate the GUP and HUP cases, respectively. We observe that for $\omega_q=-1$ the energy-matter density is independent of horizon radius. In Fig. \ref{fig:4c} and Fig. \ref{fig:4d} we compare the HUP and GUP cases of  $\omega_q=-2/3$ and $\omega_q=-1/3$. We see that energy-matter density decreases as linear inverse and inverse square of horizon radius in the absence of GUP-correction. The presence of GUP-correction slightly increases the value of the energy-matter density function especially for the small values of radius.  

Finally, we derive the equation of state of the black hole by taking the relations between the pressure and the quintessence matter-energy density, which are given with Eqs. \eqref{Pressure} and \eqref{energymatter}, into account. Then, we express the mass in terms of the pressure and horizon radius. 
\begin{eqnarray}
M&=& \frac{r_H}{2}+\frac{ P_q r_H^3}{3\omega_q^2}.
\end{eqnarray}
According to $V=\left(\frac{\partial M}{\partial P_q}\right)$, we express the volume in terms of the horizon radius and state parameter \cite{Tharanath}:
\begin{eqnarray}
V&=& \frac{r_H^3}{3\omega_q^2}.
\end{eqnarray}
Then, we use them in Eq. \eqref{TGUP}, and we get 
\begin{eqnarray}
T&=& \frac{2\left(3 \omega_q^2 V \right)^{1/3}}{\pi\beta}\left( 1-\frac{2 P_q \left(3 \omega_q^2 V \right)^{2/3}}{\omega_q}\right) \left( 1-\sqrt{1-\frac{\beta }{4\left(3 \omega_q^2 V \right)^{2/3}}}\right).
\end{eqnarray}
For $T=1$ isotherm, we obtain the GUP-corrected equation of state of the black hole in the form of
\begin{eqnarray}
P_q&=& \frac{\omega_q}{2\left(3 \omega_q^2 V \right)^{2/3}}\left[ 1-\frac{\pi\beta}{2\left(3 \omega_q^2 V \right)^{1/3}}\frac{1}{ 1-\sqrt{1-\frac{\beta }{4\left(3 \omega_q^2 V \right)^{2/3}}}}\right],  \label{PVGUP}
\end{eqnarray}
which reduces to
\begin{eqnarray}
P_q&=& \frac{\omega_q}{2\left(3 \omega_q^2 V \right)^{2/3}}\left[ 1-4\pi\left(3 \omega_q^2 V \right)^{1/3}\right] \label{PVHUP}
\end{eqnarray}
in the HUP limit. In order to obtain a real-valued pressure in the GUP case the following constraint arises 
\begin{eqnarray}
 V  &\geq & \frac{1}{3 \omega_q^2}\left(\frac{\beta}{4}\right)^\frac{3}{2}. 
\end{eqnarray}

\begin{figure}[htbp]
\centering
    \begin{subfigure}{0.45\linewidth}
\includegraphics[width=\linewidth]{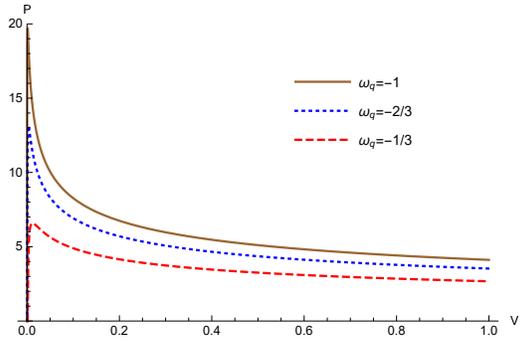} 
    \caption{HUP case for $\alpha=0$.}
\label{fig:5a}
    \end{subfigure}\hfill
    \begin{subfigure}{0.45\linewidth}
\includegraphics[width=\linewidth]{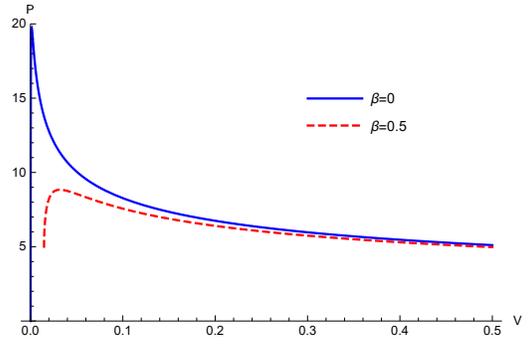}
    \caption{$w_q=-1$ case.}
\label{fig:5b}
    \end{subfigure}
    \begin{subfigure}{0.45\linewidth}
\includegraphics[width=\linewidth]{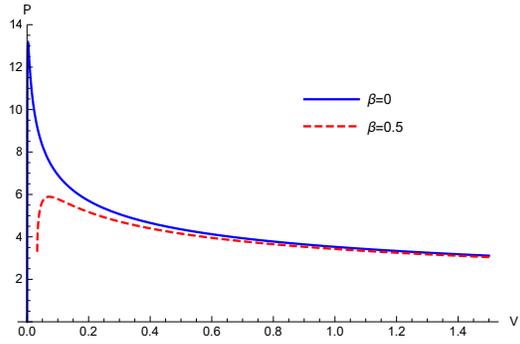}
    \caption{$w_q=-2/3$ case.}
\label{fig:5c}
    \end{subfigure}\hfill
    \begin{subfigure}{0.45\linewidth}
\includegraphics[width=\linewidth]{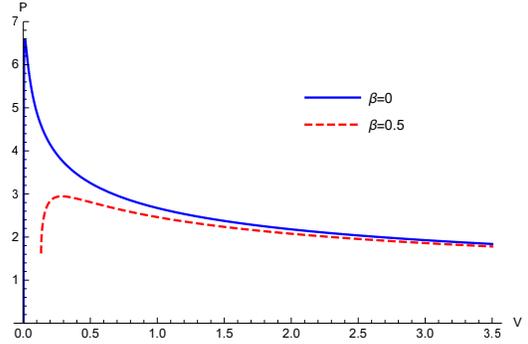}
    \caption{$w_q=-1/3$ case.}
\label{fig:5d}
    \end{subfigure}
\caption{Pressure versus volume for $\alpha=0.5$.}
    \label{fig:5}
    \end{figure}

Finally, we summarize these results according to the particular quintessence state parameter values as we have done above.
\begin{itemize}
    \item For $\omega_q=-1$, Eq. \eqref{PVGUP} becomes
\begin{eqnarray}
P_q&=& -\frac{1}{2\left(3 V \right)^{2/3}}\left[ 1-\frac{\pi\beta}{2\left(3 V \right)^{1/3}}\frac{1}{ 1-\sqrt{1-\frac{\beta }{4\left(3  V \right)^{2/3}}}}\right],  \label{PVGUP-1}
\end{eqnarray}
and in the HUP limit it reduces to
\begin{eqnarray}
P_q&=&  -\frac{1}{2\left(3 V \right)^{2/3}}\left[ 1-4\pi\left(3 V \right)^{1/3}\right].  \label{PVHUP-1}
\end{eqnarray}

    \item For $\omega_q=-2/3$, Eq. \eqref{PVGUP} turns to the form of
\begin{eqnarray}
P_q&=& -\frac{1}{3\left(\frac{4V}{3} \right)^{2/3}}\left[ 1-\frac{\pi\beta}{2\left(\frac{4V}{3}\right)^{1/3}}\frac{1}{ 1-\sqrt{1-\frac{\beta }{4\left(\frac{4V}{3} \right)^{2/3}}}}\right],  \label{PVGUP-23}
\end{eqnarray}
and in the HUP limit it gives
\begin{eqnarray}
P_q&=&  -\frac{1}{3\left(\frac{4V}{3} \right)^{2/3}}\left[ 1-4\pi\left(\frac{4V}{3} \right)^{1/3}\right].  \label{PVHUP-23}
\end{eqnarray}

    \item For $\omega_q=-1/3$, Eq. \eqref{PVGUP} becomes
\begin{eqnarray}
P_q&=& -\frac{1}{6\left(\frac{V}{3} \right)^{2/3}}\left[ 1-\frac{\pi\beta}{2\left(\frac{V}{3}\right)^{1/3}}\frac{1}{ 1-\sqrt{1-\frac{\beta }{4\left(\frac{4}{3} \right)^{2/3}}}}\right],  \label{PVGUP-13}
\end{eqnarray}
and in the HUP limit we find
\begin{eqnarray}
P_q&=&  -\frac{1}{6\left(\frac{V}{3} \right)^{2/3}}\left[ 1-4\pi\left(\frac{V}{3} \right)^{1/3}\right].  \label{PVHUP-13}
\end{eqnarray}
\end{itemize}
In Fig. \ref{fig:5}, we plot the P-V isotherm. In Fig. \ref{fig:5a}, we demonstrate the HUP case results. We observe that while the volume increases the pressure shows a slower decrease for higher quintessence state parameter. In Fig. \ref{fig:5a}, Fig. \ref{fig:5b} and Fig.\ref{fig:5c}, we demonstrate the effects of the GUP-correction on the equation of state of the case for $\omega_q=-1$, $\omega_q=-2/3$, and $\omega_q=-1/3$, respectively. We find that with the GUP-corrections the pressure of the black hole achieves smaller values. Moreover, we observe that the volume of the black hole is bounded from below in the GUP case as predicted above. On the other hand the GUP-corrections lose their effect in the higher volume values.  

\section{Conclusion}
In this manuscript, we examine the effects of the generalized uncertainty principle on the thermodynamics of the Schwarzschild black hole surrounded by quintessence. We find that two conditions arise which {bound} the event horizon radius from above and below. We observe that the upper bound depends on the quintessence while the lower bound depends on the deformed algebra. We see that these constraints {are effective by setting limits on temperature values.} Then, we {study} the GUP-corrected heat capacity and entropy functions of the black hole by considering the laws of thermodynamics.  We observe a non-zero remnant temperature in the presence of the deformed algebra. Moreover, we find logarithmic terms in the GUP-corrected entropy. Then, we examine the equation of state of the black hole. We find that the quantum effects slightly increase the energy-matter density function. Finally, we investigate the pressure-volume isotherm of the black hole. We observe the effects of deformed algebra on pressure more dominantly in smaller volumes. We compare the ensuing results through graphical methods in the context of the deformed and undeformed Heisenberg algebra. Moreover, we demonstrate the effect of the GUP correction on the thermal properties of the black hole for three particular values of the quintessence state parameter and make comparison among them.

{
\section*{Acknowledgements}
The authors thank the anonymous reviewer for his/her helpful and constructive comments.}

\end{document}